\documentstyle[epsfig]{mn}

\title[Remnant velocity distributions]{The line of sight velocity
distributions of simulated merger remnants}

\author[G.J. Bendo and J.E. Barnes]{George J. Bendo and Joshua E. Barnes\\
        Institute for Astronomy, 2680 Woodlawn Dr., Honolulu, HI 96822, USA}

\begin{document}

\maketitle

\begin{abstract}
We use Gauss-Hermite functions to study the line of sight velocity
distributions in simulated merger remnants.  Our sample contains
sixteen remnants; eight produced by mergers between disk galaxies of
equal mass and eight produced by mergers between disk galaxies with
mass ratios of 3:1.  The equal-mass mergers display a wide range of
kinematic features, including counterrotation at large radii,
orthogonally rotating cores, and misaligned rotational axes.  Most of
the unequal-mass remnants exhibit fairly regular disk-like kinematics,
although two have kinematics more typical of the equal-mass remnants.
Our results may be compared to observations of early type objects,
including ellipticals with misaligned kinematic axes, counterrotating
systems, and S0 galaxies.
\end{abstract}


\section{Introduction}

Kinematic studies of early-type galaxies have revealed a remarkable
variety of interesting behavior; some galaxies have rotation axes
``misaligned'' with respect to their minor axes (Franx, Illingworth,
\& de Zeeuw 1991), while in others the inner regions counterrotate
with respect to the rest of the galaxy (Statler, Smecker-Hane, \&
Cecil 1996; Bender \& Surma 1992; van der Marel \& Franx 1993).  Such
intriguing kinematics could plausibly result if these galaxies are the
end-products of disk-galaxy mergers (Toomre \& Toomre 1972), and
N-body simulations have gone some ways toward showing that mergers can
indeed produce remnants with distinctive kinematics (Hernquist \&
Barnes 1991; Barnes 1992, 1998; Balcells \& Gonz\'alez 1998).
However, other theories have been put forward for such kinematic
features, particularly in the case of counterrotation (Kormendy 1984;
Bertola, Buson, \& Zeilinger 1988).  Distinguishing between major
mergers and other explanations for distinctive kinematics in galaxies
has been especially difficult.

The projected luminosity profiles and isophotal shapes of simulated
disk galaxy mergers are reasonably good matches to those of elliptical
galaxies (eg.~Barnes 1988; Hernquist 1992, 1993; Governato, Reduzzi,
\& Rampazzo 1993; Heyl, Hernquist, \& Spergel 1994), but few workers
have investigated the projected {\it kinematics\/} of simulated merger
remnants.  Hernquist (1992, 1993) described principal-axis profiles of
projected mean velocity and velocity dispersion for several disk-disk
merger remnants, and Heyl, Hernquist, \& Spergel (1996) studied line
of sight velocity distributions for a somewhat larger sample of
objects.  These studies showed that kinematic misalignments of merger
remnants are observable, and indicated that skewness of line profiles
could provide information on the initial orientations of the merging
disks.  However, while systematically exploring different projections,
these studies were limited to equal-mass mergers, and did not examine
the structure of line profiles in detail or map velocity fields in two
dimensions.

Therefore, we studied line of sight velocity distributions for a
larger sample of simulated merger remnants.  We examined eight mergers
between disk galaxies with mass ratios of 1:1 and another eight
mergers between disk galaxies with mass ratios of 3:1.  We limited our
analysis to a single projection along the intermediate axis of each
remmant, but we complement a extensive presentation of major-axis
kinematics with detailed examinations of individual line profiles and
with two-dimensional maps of key kinematic parameters.  This work
extends the studies described above to unequal-mass mergers, clarifies
the connection between initial conditions and line profile, and
provides predictions to be compared with kinematic studies of
early-type galaxies using the next generation of integral-field
spectrometers.

The outline of this paper is as follows.  The rest of Section~1
describes the merger simulations and the methods we use to extract
line of sight velocity distributions and represent the distributions
with Gauss-Hermite parameters.  Sections~2 and~3 present the results
for the equal-mass and unequal-mass mergers, respectively.  Section~4
compares our results to observational studies and summarizes our
conclusions.

\subsection{Merger simulations}

The remnants analyzed here came from a modest survey of parabolic
encounters between model disk galaxies (Barnes 1998).  Each model had
three components: a central bulge with a shallow cusp (Hernquist
1990), an exponential/isothermal disk with constant scale height
(Freeman 1970; Spitzer 1942), and a dark halo with a constant-density
core (Dehnen 1993; Tremaine et al.~1994).  Density profiles for these
components are
\begin{eqnarray}
    \rho_{\rm b} &\propto& r^{-1} (r + a_{\rm b})^{-3} \, , \\
    \rho_{\rm d} &\propto& \exp(-R/R_{\rm d}) \,
                             {\rm sech}^2(z/z_{\rm d}) \, , \\
    \rho_{\rm h} &\propto& (r + a_{\rm h})^{-4} \, , 
\end{eqnarray}
where $r$ is spherical radius, $R$ is cylindrical radius in the disk
plane, and $z$ is distance from the disk plane.

Adopting simulation units with $G = 1$, the model used in the
equal-mass mergers has a bulge mass of $M_{\rm b} = 0.0625$, a disk
mass of $M_{\rm d} = 0.1875$, and a halo mass of $M_{\rm h} = 1$.  The
bulge scale length is $r_{\rm b} = 0.0417$, the disk scale radius and
scale height are $R_{\rm d} = 0.0833$ and $z_{\rm d} = 0.007$, and the
halo scale radius is $r_{\rm h} = 0.1$.  With these parameter choices,
the model has a half-mass radius $r_{1/2} \simeq 0.28$, and the
circular velocity and orbital period at this radius are $v_{1/2}
\simeq 1.5$ and $t_{1/2} \simeq 1.2$.  The model may be roughly scaled
to the Milky Way by equating our units of length, mass, and time to
$40 {\rm\,kpc}$, $2.2 \times 10^{11} {\rm\,M_\odot}$, and $2.5 \times
10^8 {\rm\,yr}$, respectively.

In the unequal-mass mergers, the larger model had the same parameters
as listed above, while the small model was scaled down by a factor of
$3$ in mass and $\sqrt{3}$ in radius in rough accord with the standard
luminosity-rotation velocity relation for disk galaxies.

Each experiment used a total of $131072$ particles, $65536$ assigned
to the luminous components, and $65536$ assigned to the dark halos.
The models were run with a tree code using a spatial resolution of
$\epsilon = 0.01$ and a time-step $\Delta t = 1/128$.  With these
integration parameters, total energy was conserved to within $0.5$\%
peak-to-peak.

All eight equal-mass merger simulations used the same initial orbit,
leading in each case to close ($r_{\rm p} \simeq 0.2$) parabolic
encounter.  Disk angles for these experiments are listed in
Table~\ref{eqmass-angles}; $i$ and $\omega$ are the inclination and
argument of pericenter (Toomre \& Toomre 1972), while the subscripts
$1$ and $2$ label the two disks.  After merging, remnants were evolved
for several more dynamical times before being analyzed.

\begin{table}
\caption{Initial disk angles for equal-mass merger simulations.}
\label{eqmass-angles}
\begin{tabular}{lrrrr}
ID & $i_1$ & $\omega_1$ & $i_2$ & $\omega_2$ \\
\noalign{\smallskip}
A  &     0 &            &    71 &         30 \\
B  &  -109 &         90 &    71 &         90 \\
C  &  -109 &        -30 &    71 &        -30 \\
D  &  -109 &         30 &   180 &            \\
E  &     0 &            &    71 &         90 \\
F  &  -109 &        -30 &    71 &         30 \\
G  &  -109 &         30 &    71 &        -30 \\
H  &  -109 &         90 &   180 &            \\
\end{tabular}
\end{table}

The eight unequal-mass merger simulations generalize the equal-mass
simulations A, B, C, and D by allowing the mass of either galaxy to
vary by a factor of three.  Like their equal-mass counterparts, these
experiments adopted a parabolic initial orbit with pericentric
separation $r_p = 0.2$.  Table~\ref{uneqmass-angles} lists the
inclinations and pericentric arguments for each simulation; here $i_1$
and $\omega_1$ are the angles for the larger disk, while $i_2$ and
$\omega_2$ are the angles for its smaller companion.

\begin{table}
\caption{Initial disk angles for unequal-mass merger simulations.}
\label{uneqmass-angles}
\begin{tabular}{lrrrr}
ID & $i_1$ & $\omega_1$ & $i_2$ & $\omega_2$ \\
\noalign{\smallskip}
A$_1$&     0 &            &    71 &         30 \\
A$_2$&    71 &         30 &     0 &            \\
B$_1$&  -109 &         90 &    71 &         90 \\
B$_2$&    71 &         90 &  -109 &         90 \\
C$_1$&  -109 &        -30 &    71 &        -30 \\
C$_2$&    71 &        -30 &  -109 &        -30 \\
D$_1$&   180 &            &  -109 &         30 \\
D$_2$&  -109 &         30 &   180 &            \\
\end{tabular}
\end{table}

\begin{figure}
\begin{center}
\epsfig{figure=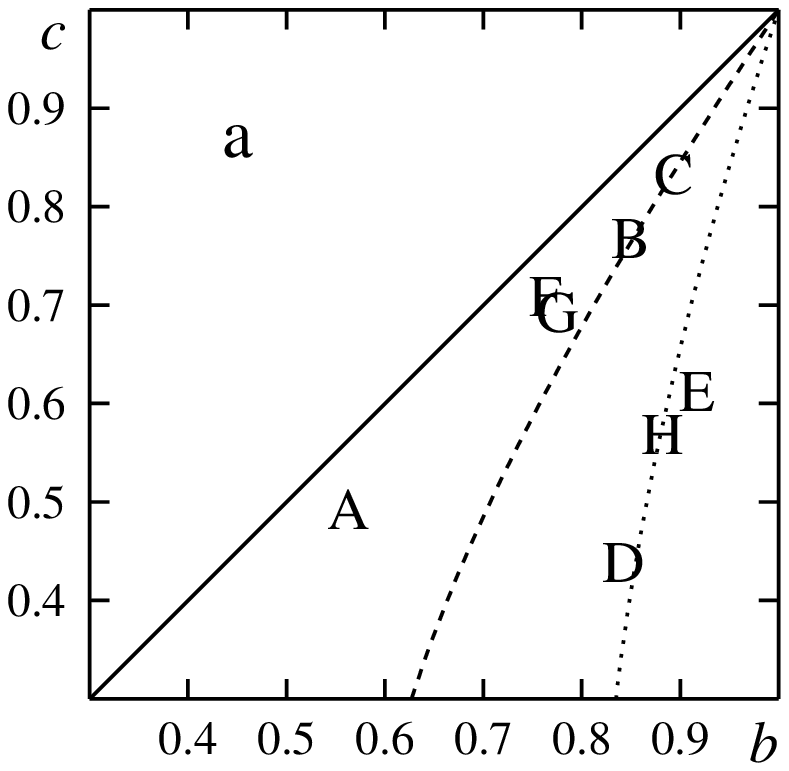, width=2in}
\end{center}
\begin{center}
\epsfig{figure=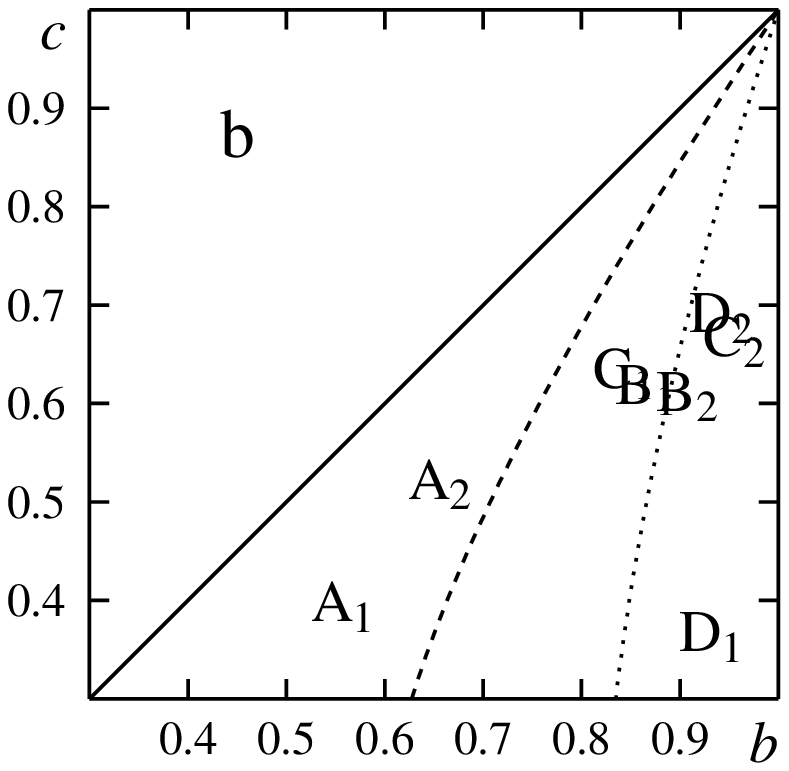, width=2in}
\end{center}
\caption{Axial ratios for remnants of (a) equal-mass mergers, and (b)
unequal-mass mergers.  Solid, dashed, and dotted lines are contours of
triaxiality $T = 1$, $\frac{2}{3}$, and $\frac{1}{3}$, respectively,
where $T \equiv (1 - b^2)/(1 - c^2)$.}
\label{shapes}
\end{figure}

Some salient properties of these merger remnants are summarized here;
for a more detailed discussion, see Barnes (1998).  All sixteen
remnants are ellipsoidal objects with luminosity profiles generally
following a de~Vaucouleurs law.  The projected half-light radii
$R_{\rm e}$ of the equal-mass remnants range from $0.133$ to $0.157$,
while for the unequal-mass remnants $R_{\rm e}$ ranges from $0.099$ to
$0.123$.  Fig.~\ref{shapes} shows axial ratios determined from the
inertia tensor for the most tightly-bound half of the luminous
particles in each object.  On the whole, the remnants of equal-mass
mergers are triaxial or prolate, while those produced by unequal-mass
mergers tend to be more oblate.

\subsection{Gauss-Hermite analysis}

For all sixteen simulated remnants we extracted five frames, each
containing $65536$ luminous particles -- that is, particles from the
bulges and disks of the progenitor galaxies.  Each set of five frames
is equally-spaced over a total of $0.5$ time units; this interval is
long enough that individual particles are sampled at effectively
random orbital phases, but not long enough for the remnant to undergo
significant evolution.  We shifted each frame to place the potential
minimum at the origin and rotated it to diagonalize the moment of
inertia tensor for all particles with a potential less than $0.8$
times the minimum potential.  In what follows, we use $X$, $Y$, and
$Z$ for the major, intermediate, and minor axes of the remnants.

To measure line of sight velocity distributions as a function of
position along a given axis, we created a two dimensional grid, with
one dimension representing position and the other dimension
representing velocity, thus simulating a slit in a spectrometer.
Typically, we placed the slit along the major ($X$) axis and projected
along the intermediate ($Y$) axis, although other options were used in
preliminary investigations.  The width of the slit was set at $0.03$,
which is roughly $20$\% of the projected half-light radius.  For each
frame from each remnant, the particles falling within the slit were
binned in position and velocity.  The grid spacing along the slit was
set to a minimum of $0.02$ and increased as necessary to keep the
total number of particles falling within the range above a minimum.
The width of the velocity bins was set to a fixed value of $0.2$,
spanning the velocity range $|v| \le 4$ with $40$ bins.

To map the line of sight velocity distributions across the plane of
the sky, we used a generalization of the above procedure.  However,
two adjustable bin dimensions were created instead of one along the
given slit.

After binning the data, the velocity distribution at each location was
fit with a parameterized Gauss-Hermite series (van der Marel \& Franx
1993).  The value of each parameter was determined by combining the
five frames and performing a least-squares fit; uncertainties were
estimated by comparing fits of individual frames.  Gauss-Hermite
functions are modified Gaussians with additional skewness and kurtosis
parameters; they provide an effective way to parameterize the
moderately non-Gaussian distributions which arise in systems which
have undergone incomplete violent relaxation.  The formula for the
fitting function is
\begin{eqnarray}
    P(v) &=& \gamma \frac{\alpha(w)}{\sigma} [1 + h_3 H_3(w) + h_4 H_4(w)] \, ,
\end{eqnarray}
where $w \equiv (v - v_0) / \sigma$ and
\begin{eqnarray}
    \alpha(w) &\equiv& \frac{1}{\sqrt{2 \pi}} e^{-w^2/2} \, , \\
    H_3(w) &\equiv& \frac{1}{\sqrt{6}}(2 \sqrt{2} w^3 - 3 \sqrt{2} w) \, , \\
    H_4(w) &\equiv& \frac{1}{\sqrt{24}}(4 w^4 - 12 w^2 + 3) \, .
\end{eqnarray}
This function has five parameters: $\gamma$, $v_0$, $\sigma$, $h_3$,
and $h_4$.  The normalization factor $\gamma$ has little physical
significance in our study.  The mean velocity $v_0$ and velocity
dispersion $\sigma$ have dimensions of velocity, while the $h_3$ and
$h_4$ parameters represent the skewness and kurtosis of the velocity
distribution and are dimensionless.  When $h_3 = h_4 = 0$, the
Gauss-Hermite series produces a normal Gaussian profile.  When $h_3$
has the same sign as $v_0$ the distribution's leading wing is broad
and the trailing wing is narrow, while when $h_3$ and $v_0$ have
opposite signs the trailing wing is broad and the leading wing is
narrow.  When $h_4 > 0$, the distribution has a narrow peak with broad
wings, and when $h_4 < 0$, the distribution has a broad peak with
narrow wings.

\subsection{Orbit classification}

In order to discover which orbital families are responsible for
various features in the velocity distributions, we assigned each
particle to an orbit family using the algorithm described in Fulton \&
Barnes (submitted).  This algorithm follows each particle for a number
of radial periods and classifies its orbit by examining the sequence
of principal plane crossings.  To save time and slightly reduce the
effects of discreteness, we calculated the trajectories using a
quadrupole-order expansion of the gravitational field (White 1983).
For the present purpose all ``boxlet'' orbits were counted as boxes;
thus the major orbital families recognized here are Z-tubes, which
rotate about the minor axis, X-tubes, which rotate about the major
axis, and boxes, which do not rotate.

\section{Equal-Mass Mergers}

A merger of comparable-mass galaxies usually eliminates many of the
initial characteristics of both galaxies in the formation of the new
galaxy.  The resulting remnants are supported partly by rotation and
may sometimes be flattened, but the overall structure of the galaxies
as well as the dynamics are radically changed.  (This is in contrast
to the 3:1 mergers, where the disk of the larger galaxy often survives
the merger.)  Furthermore, the dynamics of individual 1:1 mergers
produced with different initial parameters vary greatly.

\subsection{Typical 1:1 merger parameter curves}

\begin{figure}
\begin{center}
\epsfig{figure=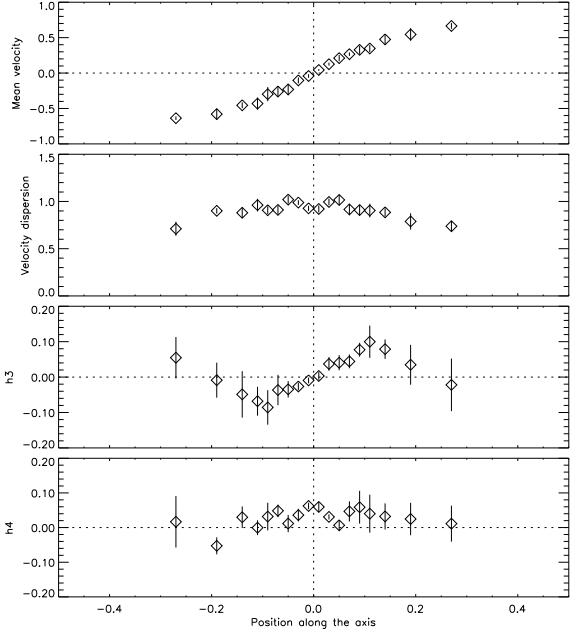,width=3.25in}
\end{center}
\caption{Parameters for the line of sight velocity distribution as
functions of position along the major axis for remnant~E, a fairly
simple 1:1 merger.}
\label{paramE}
\end{figure}

\begin{figure}
\begin{center}
\epsfig{figure=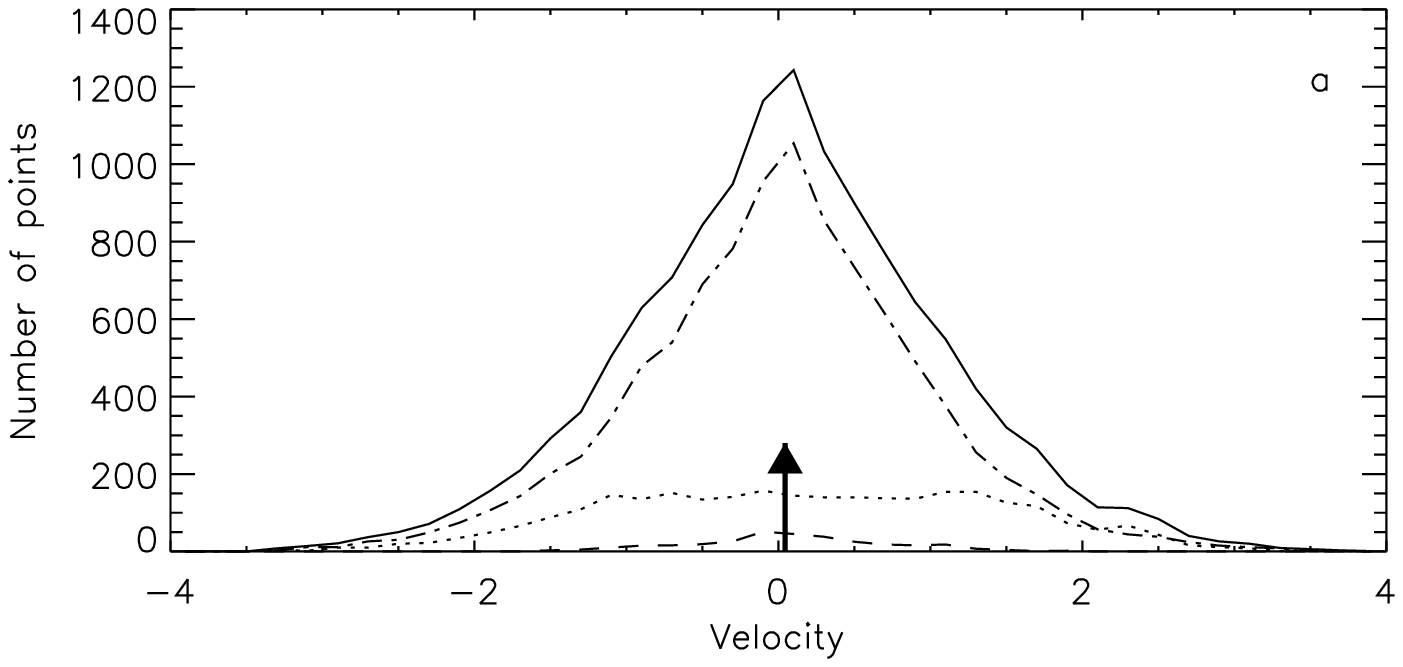,width=3.25in}
\epsfig{figure=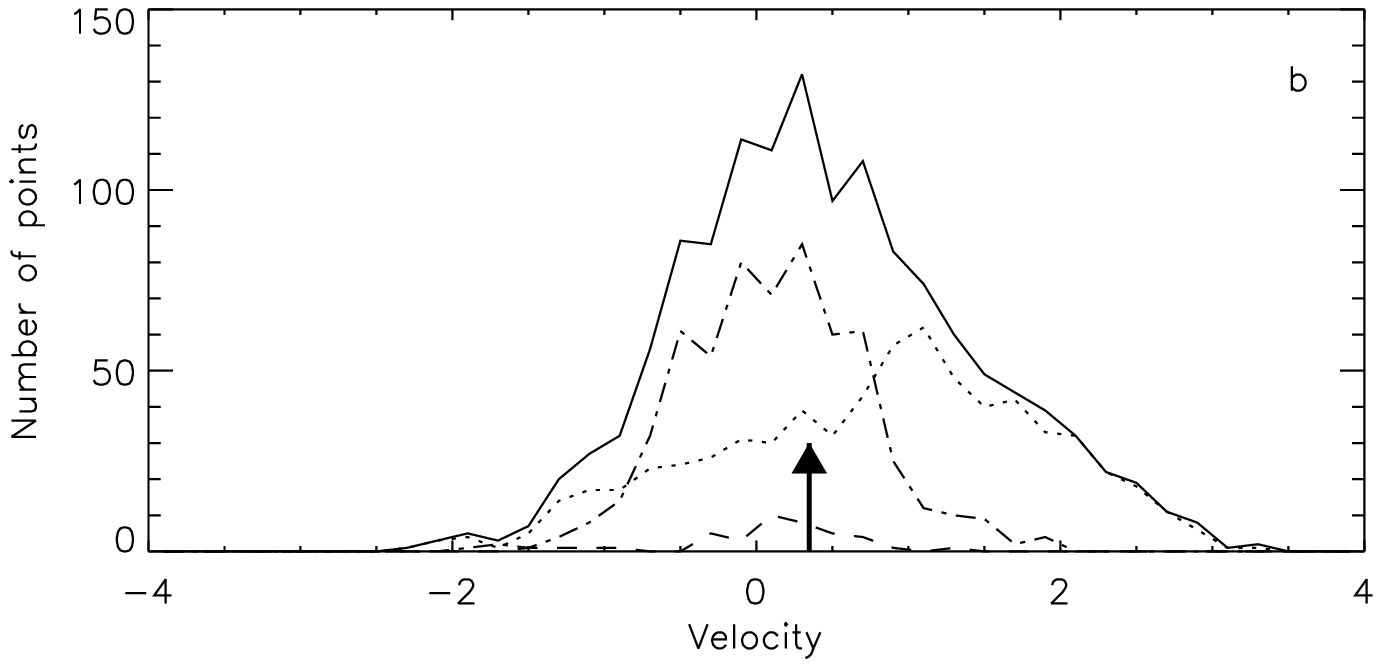,width=3.25in}
\end{center}
\caption{Line of sight velocity distributions for remnant~E on the
major axis (a) at $X = 0.01$ and (b) at $X = 0.11$ from the center.
The lines represent Z-tubes (dotted), X-tubes (dashed), and box orbits
(dash-dot), as well as all orbits combined (solid).  The vertical
arrows show mean velocities of the total distributions.  Note that
(a), which is close to the center, has relatively broad wings and a
narrow peak, while (b), which is further from the center, has a skewed
distribution, with a broad leading wing and a narrow trailing wing.}
\label{lsvdE}
\end{figure}

Fig.~\ref{paramE} plots the Gauss-Hermite parameters as functions of
position along the major axis for remnant~E.  This nearly oblate and
rapidly rotating object was produced by a direct encounter between
disks with inclinations of $i_1 = 0$ and $i_2 = 71$; it has a fairly
simple structure which contrasts the more complex cases described
below.  Fig.~\ref{lsvdE} shows examples of line of sight velocity
distributions at two different places in the galaxy.  These
distributions are shown for all the particles, and for particles
sorted by orbital family (Z-tubes, X-tubes, and box orbits).

Within one about effective radius, the mean velocity shows a roughly
linear trend with position, while at larger radii the velocity profile
rises more gradually and may start to level off.  This shape is seen
in all but one the remnants studied here, though the profile amplitude
and the radius at which it levels off varies from one remnant to
another.  The exception, in which the outskirts of the remnant
counterrotate with respect to the rest, will be described shortly.
None of the 1:1 remnants attain rotation velocities comparable to
their circular velocities, implying that rotation plays a relatively
minor role in supporting these objects.

The velocity dispersion profile in Fig.~\ref{paramE} climbs from a
local minimum at the center to a gentle peak on either side, and then
falls off slowly at greater distances from the origin.  Similar
profiles are seen in all of the 1:1 remnants; this uniformity may be
understood from the Jeans equations, since all of these moderately
anisotropic remnants have similar density profiles.  The central
regions have a large percentage of particles from the Hernquist-model
bulges of the progenitor galaxies.  These particles still follow $\sim
r^{-1}$ density profiles at small $r$ and their dispersion therefore
scales as $\sim r^{1/2}$, producing the central minima noted above.
At larger radii the rather gradual fall-off in $\sigma$ may reflect
the increasing contribution of dark matter, which dominates the mass
budget beyond about one effective radius.

In this as in most 1:1 remnants, the $h_3$ parameter is has the same
sign as the mean velocity $v_0$; as Fig.~\ref{lsvdE}b shows, the
velocity profile has broad leading and narrow trailing wings.  This
asymmetric profile arises through a combination of box orbits and
Z-tube orbits.  The former, which have a symmetric and rather narrow
velocity distribution, effectively localize the peak of the profile,
while the latter, which have a wide distribution with nonzero mean,
populate the broad leading wing.  Further from the center the trend of
$h_3$ with position is reversed, and the outermost points are
consistent with $h_3 = 0$; this may occur because the outer regions of
this rather oblate remnant are almost exclusively populated by Z-tube
orbits.

The $h_4$ profile in Fig.~\ref{paramE} shows a significant peak at
the center, falls off rapidly at slightly larger radii, then appears
to increase again before becoming statistically consistent with $h_4 =
0$ at the outer points plotted.  All the remnants in our sample have
$h_4 > 0$ at small radii and more nearly Gaussian distributions at
large radii.  However, there are large variations from remnant to
remnant, so it's not clear if the results presented for remnant~E
should be considered typical.  As Fig.~\ref{lsvdE}a shows, the
distinctly triangular shape of the velocity distribution near the
center is almost entirely due to particles on box orbits.  Further
from the center the fraction of box orbits declines and the velocity
distribution of the box orbits becomes less triangular; both of these
trends tend to reduce the measured values of $h_4$.

\begin{figure}
\begin{center}
\epsfig{figure=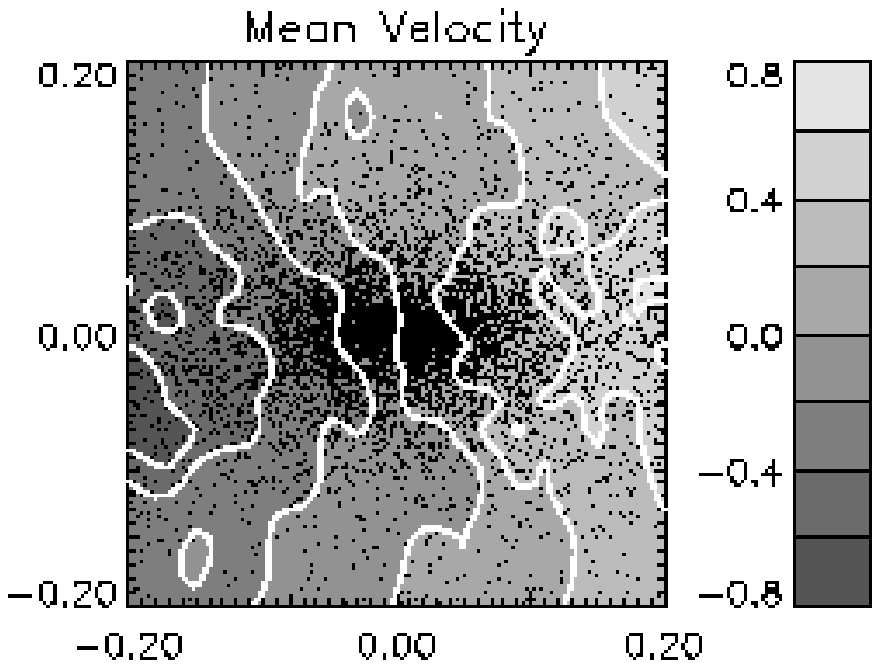,width=2.75in}
\end{center}
\begin{center}
\epsfig{figure=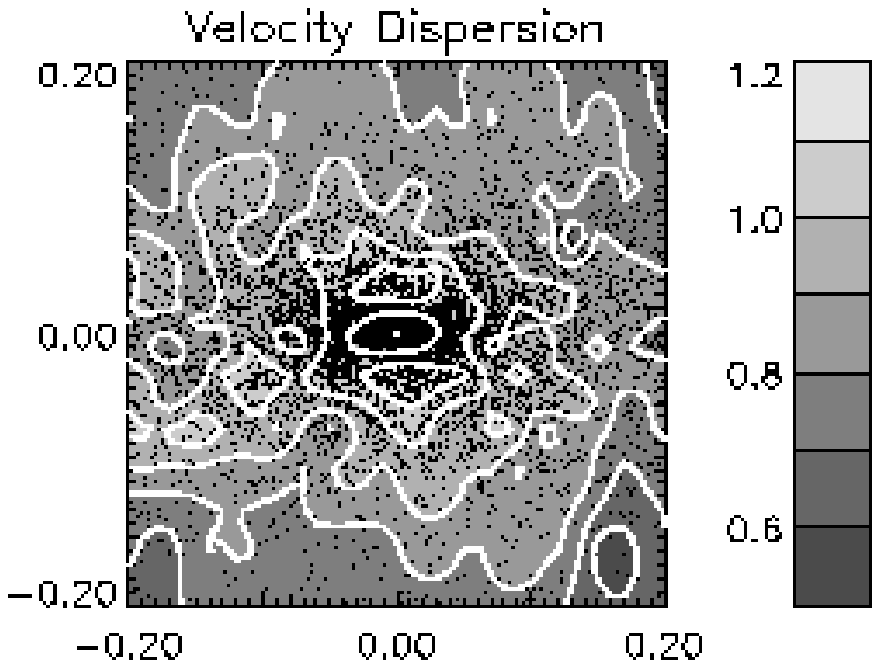,width=2.75in}
\end{center}
\caption{Contour plots of the first two Gauss-Hermite parameters for
remnant~E.  The levels are indicated to the right of each map.  A
sample of particles from one frame are plotted as points.  The
effective radius is $0.134$ for this remnant.  Errors in mean velocity
and velocity dispersion are fairly small over most of the area shown;
see Fig.~\ref{paramE}.}
\label{mapE}
\end{figure}

Having examined the behavior of the parameters on the major axis, we
briefly describe two-dimensional maps of the mean velocity and
velocity dispersion in Fig.~\ref{mapE}.  Mean rotation velocities
are highest in the equatorial plane, while elsewhere a roughly
cylindrical rotation pattern is seen.  The zero velocity contour is
slightly tilted with respect to the $Z$ axis, indicating a modest
amount of rotational misalignment at larger radii.  If the galaxy only
contained Z-tube orbits, the contours would run parallel to the $Z$
axis; the net streaming of particles in X-tube orbits cause the slant
and misaligns the rotational axes.

On the whole, the velocity dispersion falls off rather more rapidly
away from the major axis than it does along the major axis; the
dispersion contours are roughly aligned with, although rounder than,
the surface density contours.  Note, however, the two closed contours
representing dispersion {\it maxima\/} directly above and below the
central minimum, and the vertical elongation of the next lower
contour.  Several more examples of this feature will be presented
shortly.

\subsection{Variety in 1:1 mergers}

In equal-mass mergers, the degree of violent relaxation depends on the
initial orientations of the progenitor disks.  Thus in contrast to the
relatively simple product of a direct encounter just described,
mergers of inclined or retrograde disks produce remnants with a wide
range of kinematic properties.  Here we describe a few examples.

\subsubsection{Counterrotation at large radii}

\begin{figure}
\begin{center}
\epsfig{figure=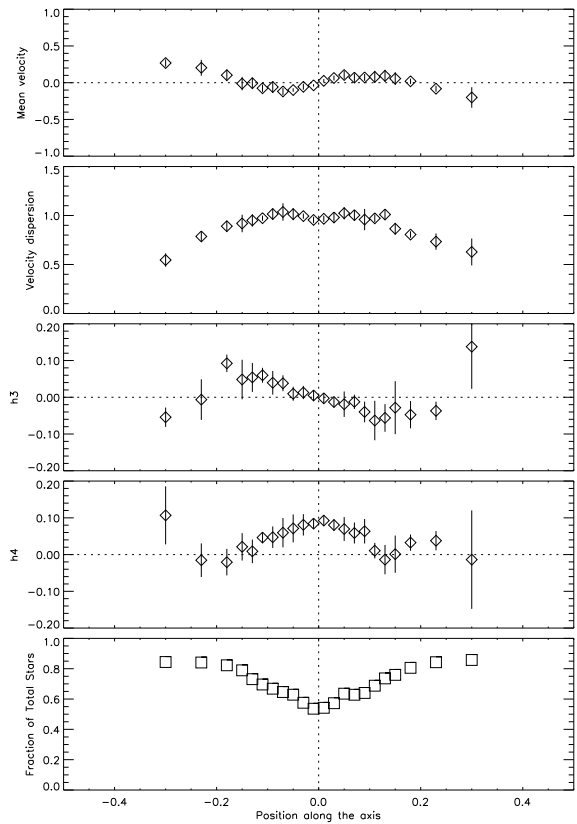,width=3.25in}
\end{center}
\caption{Parameters along the major axis for remnant~H, showing that
rotation reverses at large radii.  In addition to the Gauss-Hermite
parameters, we plot the fraction of particles from progenitor~2
against position along the major axis.  Near the center almost half of
the particles are from the first progenitor, while further out the
vast majority come from the second progenitor; this spatial
segregation accounts for the counterrotation seen at large radii.}
\label{paramH}
\end{figure}

Fig.~\ref{paramH} shows Gauss-Hermite parameters as functions of
major-axis position for remnant~H.  This object is similar in shape to
remnant~E but rotates more slowly than any other remnant in our
sample; it was produced by a retrograde encounter between disks with
inclinations of $i_1 = -109$ and $i_2 = 180$.  In the course of the
encounter, the first disk absorbed a good deal of orbital angular
momentum, producing a spheroidal structure which rotates in the same
sense as the progenitors once orbited each other.  The second disk,
which suffered an exactly retrograde passage, remained relatively thin
and retained its original sense of rotation.  This rather peculiar
combination of circumstances accounts for the counterrotation seen at
large radii.  Near the center, the virtual slit includes a substantial
fraction of particles from the first progenitor, while at larger radii
the kinematics near the major axis are dominated by particles from the
second disk.  This produces a change in the direction of mean velocity
as the distance along the major axis from the center increases.

Besides its counterrotation, this remnant has other peculiar kinematic
features.  For example, $h_3$ and $v_0$ have opposite signs over most
of the range plotted in Fig.~\ref{paramH}, implying that the velocity
profile has narrow leading and broad trailing wings; this is atypical
for an equal-mass merger remnant.  Moreover, within the effective
radius the $h_4$ parameter is quite high, indicating that the
profile's wings are relatively broad in comparison to the overall
dispersion.  These broad wings may result from the superposition of
two distinct velocity systems with widely-separated mean velocities.

\subsubsection{Major-axis rotation at small radii}

\begin{figure}
\begin{center}
\epsfig{figure=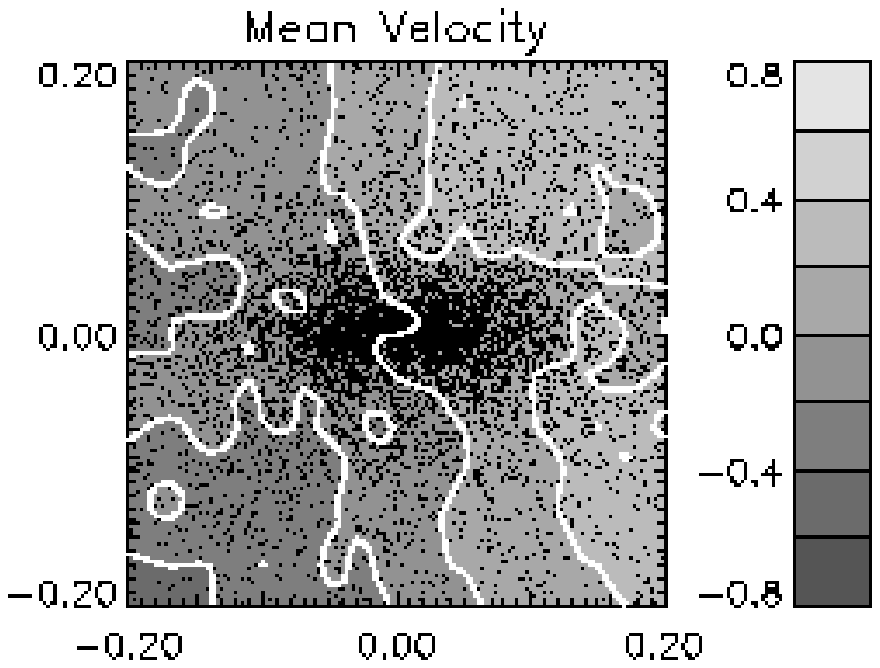,width=2.75in}
\end{center}
\begin{center}
\epsfig{figure=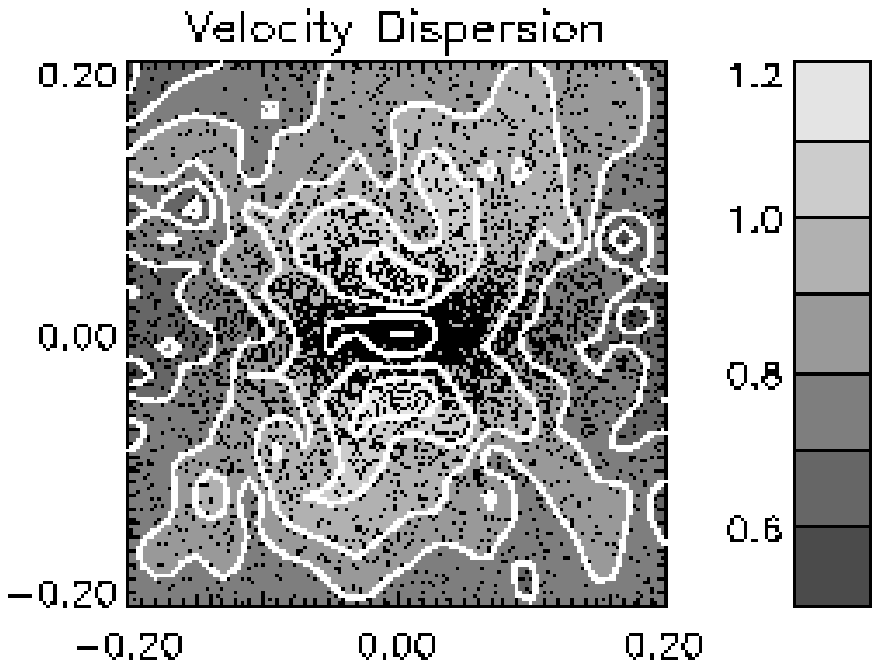,width=2.75in}
\end{center}
\caption{Contour plots of the first two Gauss-Hermite parameters for
remnant~G, in which the core rotates about the major axis.  Note the
sign change in the minor-axis rotation profile.}
\label{mapG}
\end{figure}

Fig.~\ref{mapG} presents maps of $v_0$ and $\sigma$ for remnant~G, a
slowly-rotating and relatively prolate object produced by a merger of
disks with inclinations $i_1 = -109$ and $i_2 = 71$.  This remnant has
a large population of X-tube orbits which dominate the net rotation at
small radii; the central regions thus rotate about the remnant's {\it
major\/} axis.  At larger radii the angular momentum vector is largely
determined by Z-tube orbits, favoring a more normal pattern of
minor-axis rotation.  Note that the zero velocity contour is inclined
with respect to the minor axis, indicating that X-tube orbits still
contribute to the net rotation; moreover, at large radii the X-tube
orbits rotate in the opposite direction than they do near the center.

The dispersion contours in Fig.~\ref{mapG} are elongated perpendicular
to the projected density contours, and the highest $\sigma$ values are
seen in two regions on the minor axis above and below the center of
the remnant.  An explanation for this behavior will be given shortly.

\subsubsection{Severe rotational misalignment}

\begin{figure}
\begin{center}
\epsfig{figure=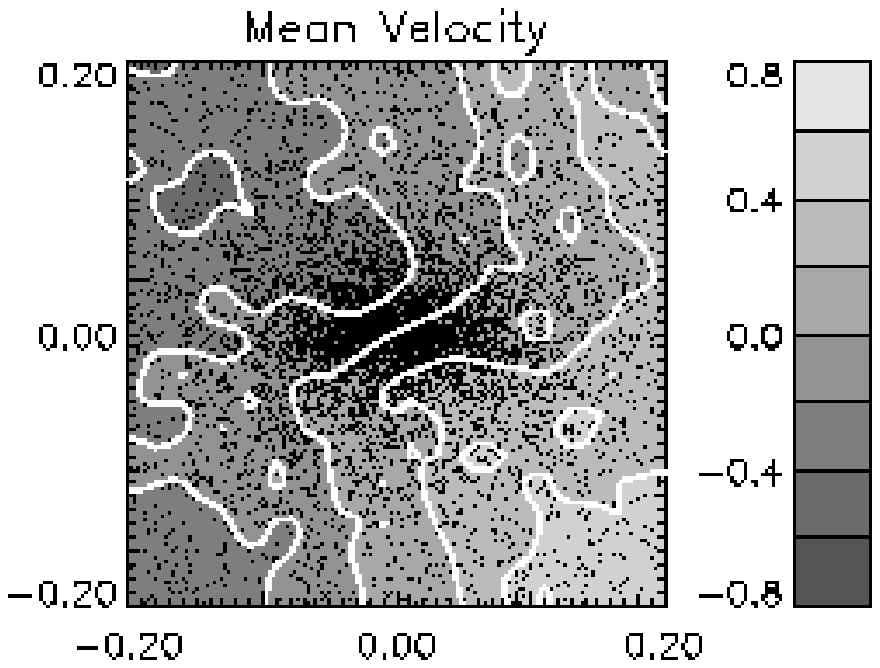,width=2.75in}
\end{center}
\begin{center}
\epsfig{figure=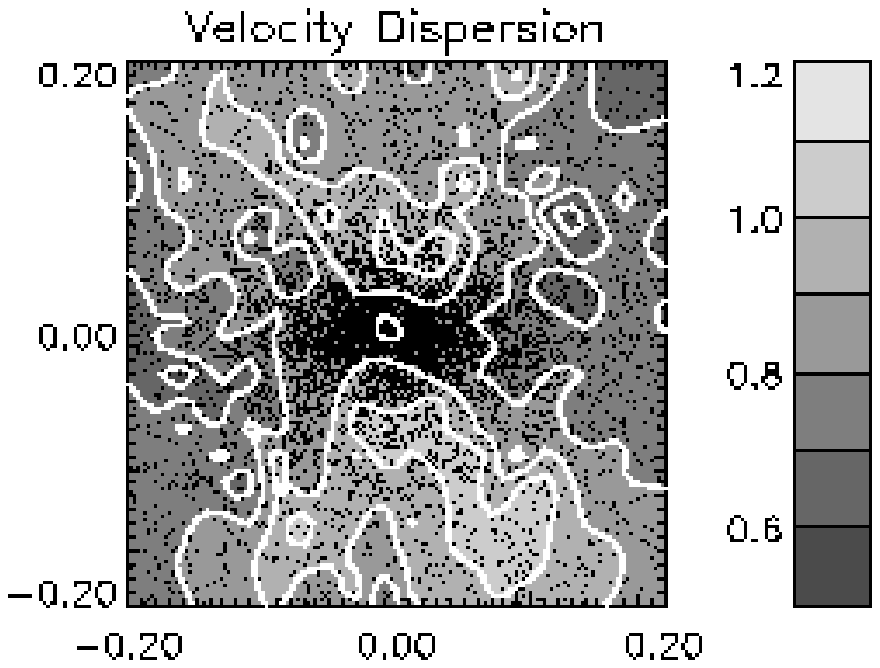,width=2.75in}
\end{center}
\begin{center}
\epsfig{figure=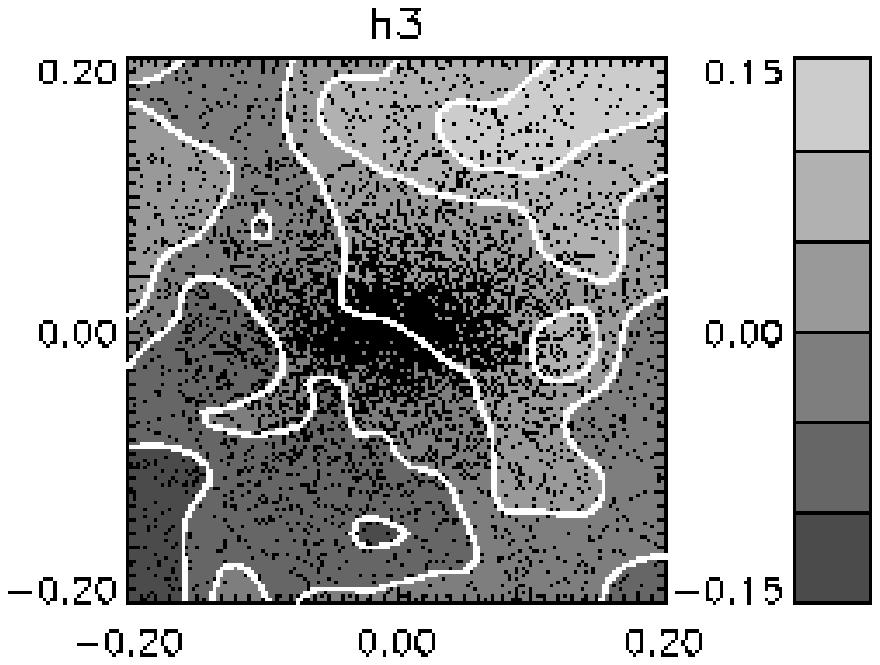,width=2.75in}
\end{center}
\begin{center}
\epsfig{figure=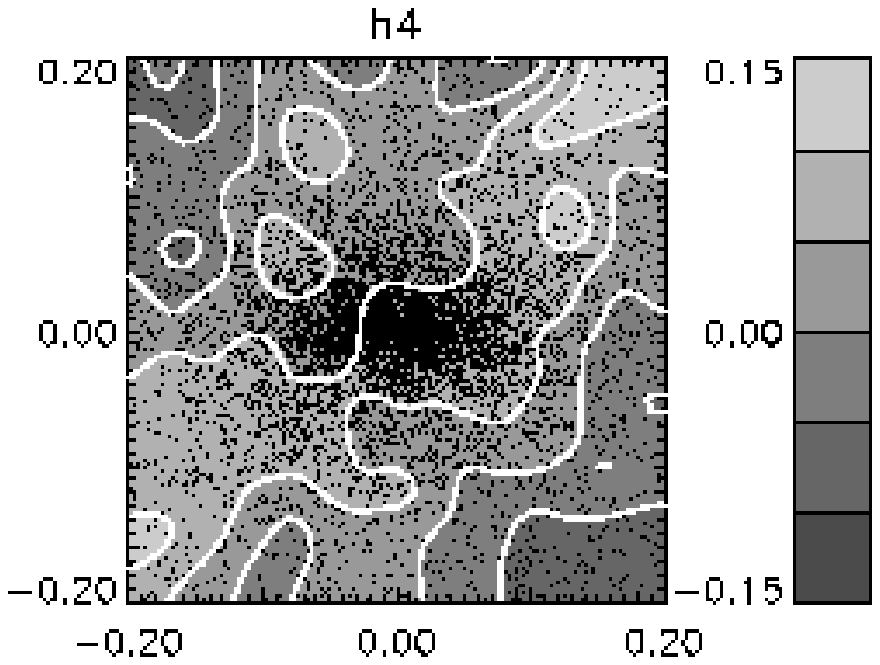,width=2.75in}
\end{center}
\caption{Contour plots of all four Gauss-Hermite parameters for
remnant~C, in which the rotational axis is severely misaligned with
respect to the minor axis.}
\label{mapC}
\end{figure}

As a last example of kinematic diversity among equal-mass mergers,
Fig.~\ref{mapC} presents maps of all four Gauss-Hermite parameters for
remnant~C.  Like remnant~G just described, this object was produced by
a merger of two inclined disks and contains a large number of X-tube
orbits.  Within the effective radius the rotation axis is severely
misaligned with the minor axis, while at larger radii the velocity
contours are more nearly parallel to the minor axis.  This occurs
because X-tube orbits make a significant contribution to the net
angular momentum at small radii, while Z-tube orbits play a larger
role at large radii.

\begin{figure}
\begin{center}
\epsfig{figure=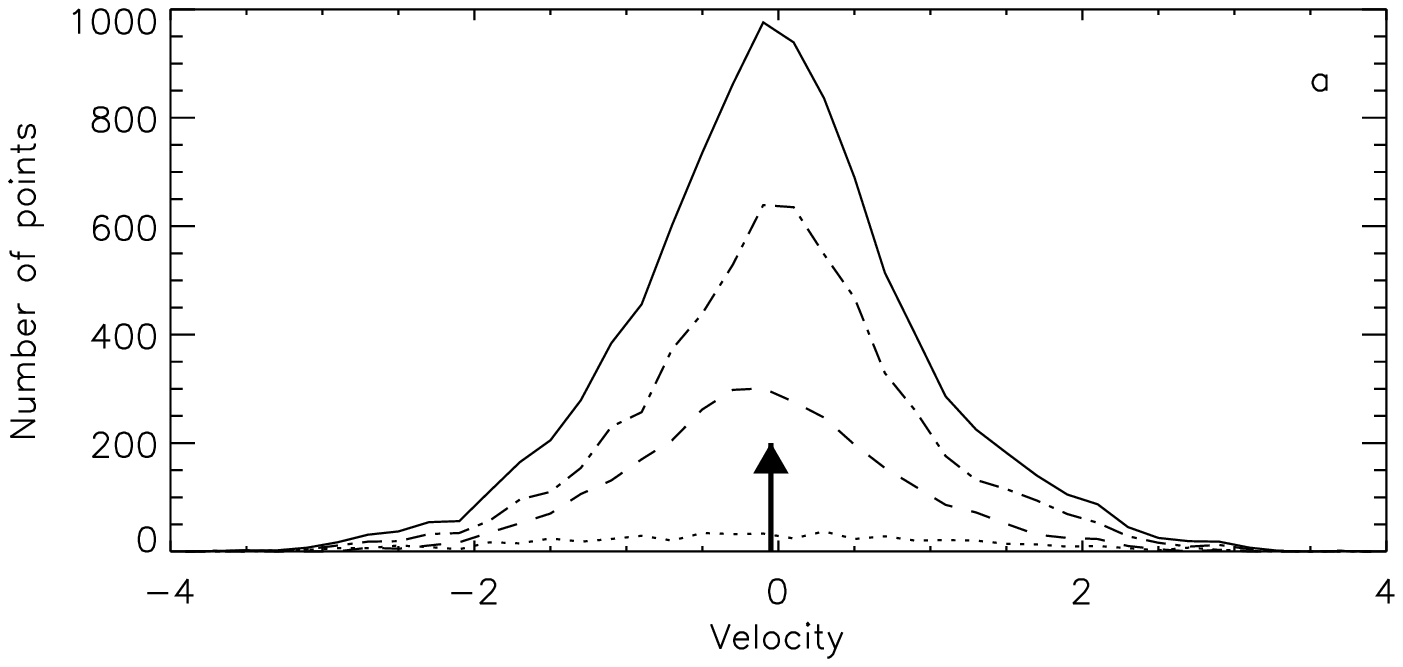,width=3.25in}
\epsfig{figure=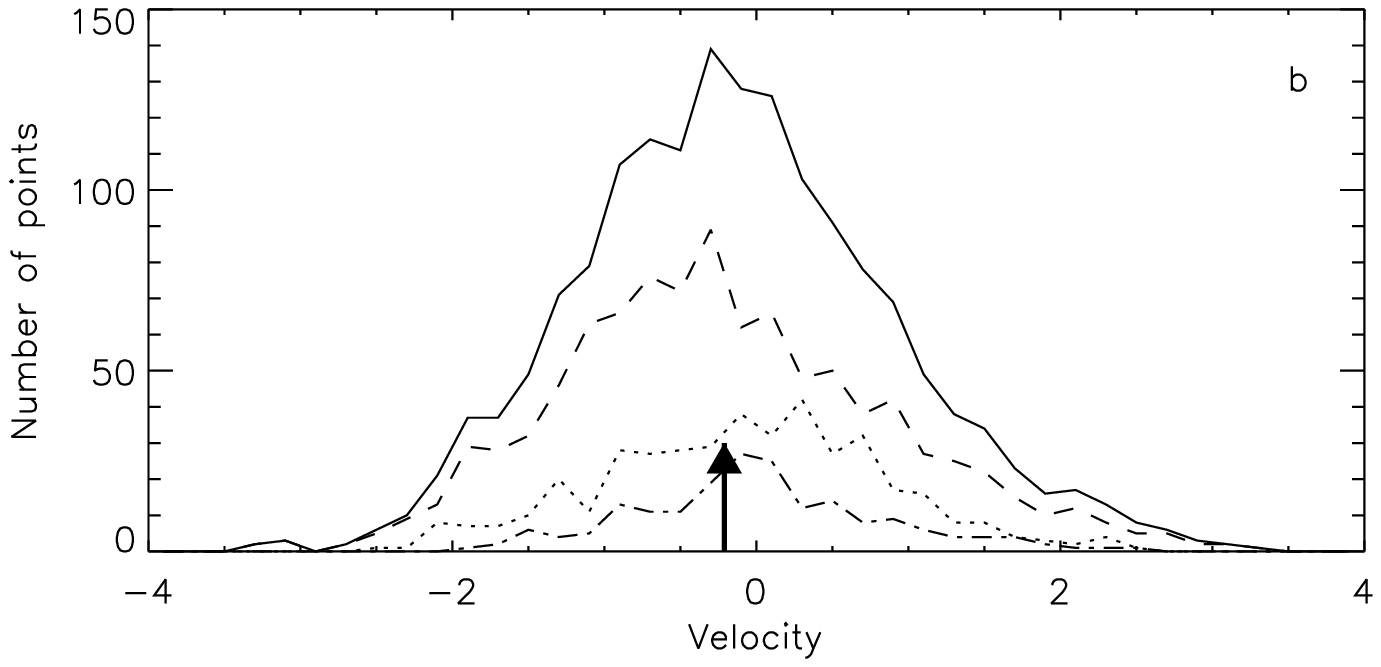,width=3.25in}
\end{center}
\caption{Line of sight velocity distributions for remnant~C on the on
the minor axis (a) at $Z = 0.01$ and (b) at $Z = 0.05$ from the
center.  As in Fig.~\ref{lsvdE}, dotted, dashed, and dash-dotted lines
represent Z-tubes, X-tubes, and boxes, respectively.  Near the center
the box orbits, which have a narrow distribution, dominate the other
populations, but above and below the major axis the X-tubes with their
broad distribution dominate.}
\label{lsvdC}
\end{figure}

Along the minor axis, the velocity dispersion peaks on either side of
the remnant's center.  Recall that such minor-axis peaks were also
seen in previous dispersion maps (Figs.~\ref{mapE} and~\ref{mapG}).
In remnants~C and~G the locations of these peaks and the general shape
of the high-dispersion regions resembles the spatial distribution of
X-tube orbits.  Such orbits encircle the waists of prolate galaxies,
and travel roughly along the $Y$ axis -- that is, towards or away from
the virtual observer -- in the regions of peak dispersion.  Further
evidence that X-tube orbits are responsible for these peaks appears in
Fig.~\ref{lsvdC}, which compares velocity profiles near the center and
near a peak of the dispersion.  The central profile, symmetric and
relatively narrow, is dominated by box orbits.  In contrast, the
profile near the peak is dominated by X-tubes; with some X-tubes
rotating in one direction and the rest in the other direction, the
velocity distribution becomes broader.  This effect is strong in
remnants~C and~G because they have large populations of X-tube orbits.

If the X-tube population was ``cold'', so that most particles stayed
close to the closed orbits which parent them, the X-tube distribution
at the peaks would be composed of two counterrotating streams and
corresponding peaks would also occur in the $h_4$ map.  No such peaks
are seen in Fig.~\ref{mapC}; instead, the $h_4$ map shows a broad
ridge of high values extending diagonally from lower left to upper
right.  The $h_3$ map, on the other hand, shows a gradient running
along the same diagonal, with low values at the lower left and high
values at the upper right.  These patterns are considerably clearer
than those seen in the other 1:1 remnants; most of the $h_3$ and $h_4$
maps we have examined seem too noisy to yield definite results.  But
we do not yet understand even the relatively simple patterns in the
present example.  For example, whereas $v_0$ and $h_3$ vary together
in the central region of remnant~E (see Fig.~\ref{paramE}), here the
contours of $v_0$ and $h_3$ are roughly orthogonal within one
effective radius.  Further modeling is needed to disentangle the roles
of different orbit families and determine the range of behavior
consistent with dynamical equilibrium.

\section{Unequal-Mass Mergers}

Mergers between galaxies of significantly different masses are less
violent than equal-mass mergers.  For sufficiently large mass ratios,
the more massive galaxy may survive essentially unscathed.  Of the
eight 3:1 mergers we examined, six have similar kinematic parameter
curves and oblate shapes.  This kinematic uniformity arises because
the larger disk basically survives the merging process.

\subsection{Oblate 3:1 mergers}

\begin{figure}
\begin{center}
\epsfig{figure=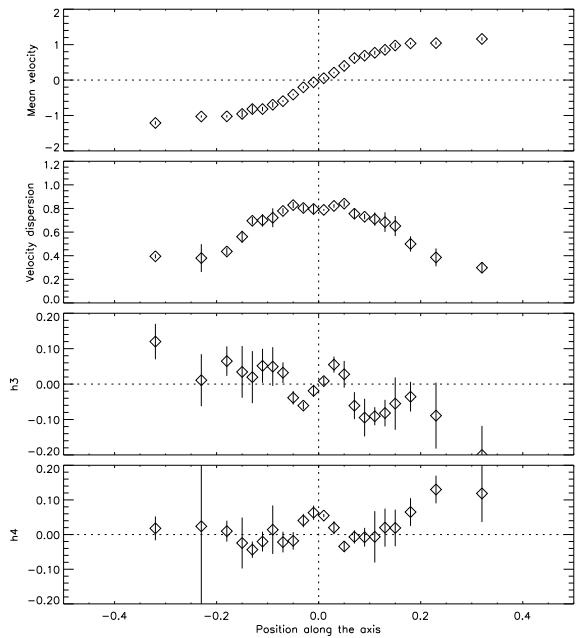,width=3.25in}
\end{center}
\caption{Parameters for the line of sight velocity distribution as
functions of position along the major axis for remnant~B$_1$, a
typical 3:1 merger.}
\label{paramB1}
\end{figure}

\begin{figure}
\begin{center}
\epsfig{figure=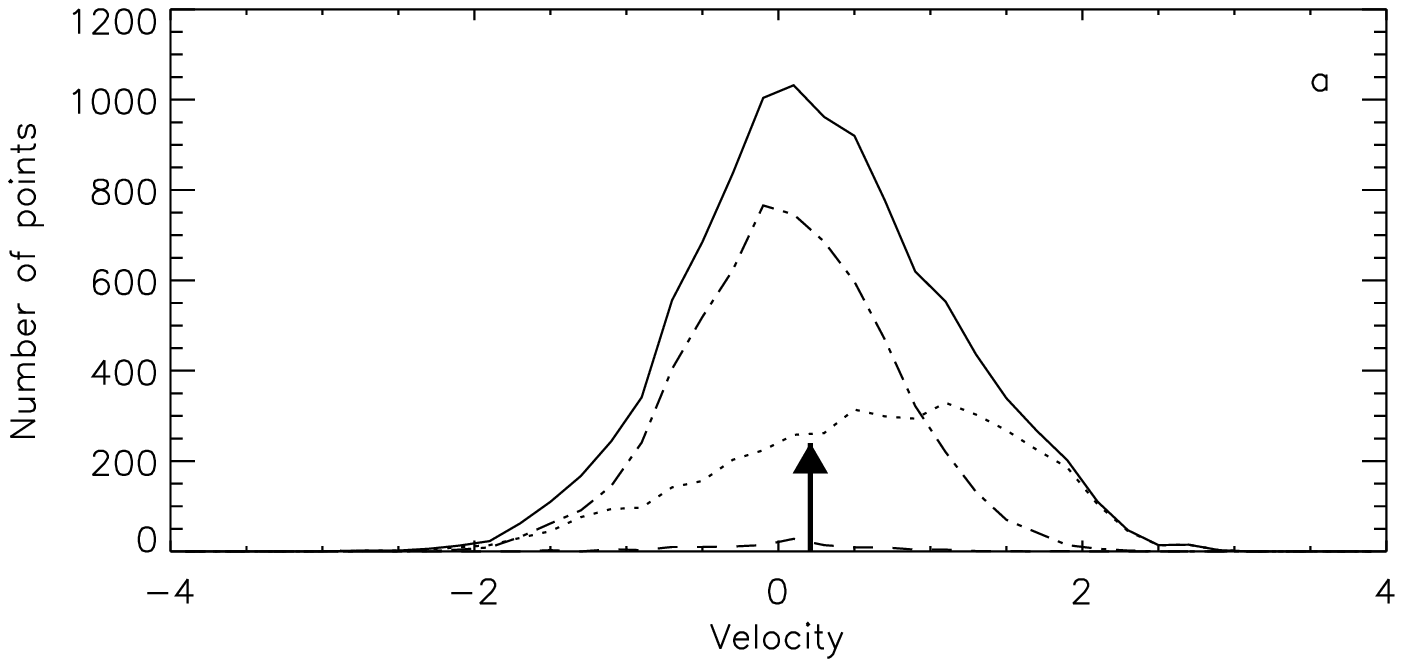,width=3.25in}
\epsfig{figure=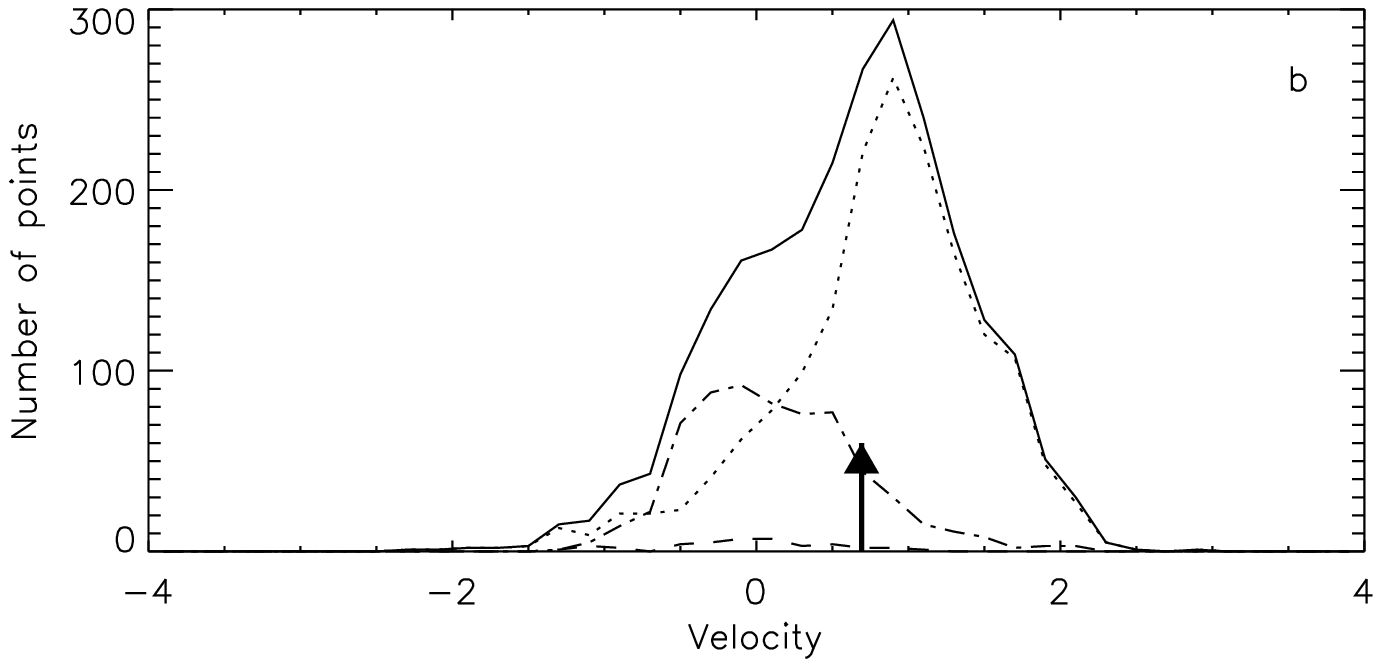,width=3.25in}
\epsfig{figure=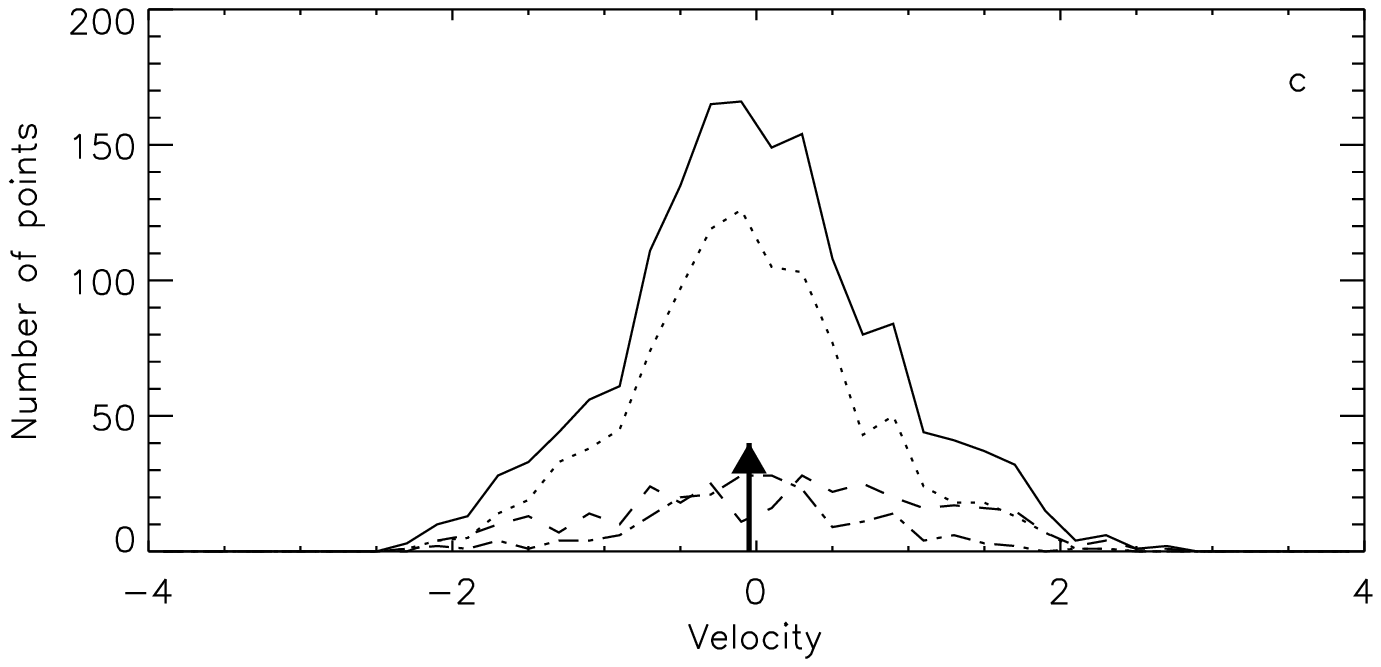,width=3.25in}
\end{center}
\caption{Line of sight velocity distributions for remnant~B$_1$ on the
major axis (a) at $X = 0.03$ and (b) at $X = 0.09$, and on the minor
axis (c) at $Z = 0.08$ from the center. As above, dotted, dashed, and
dash-dotted lines represent Z-tubes, X-tubes, and boxes,
respectively.}
\label{lsvdB1}
\end{figure}

Fig.~\ref{paramB1} plots the Gauss-Hermite parameters as functions of
position along the major axis for remnant~B$_1$.  This remnant was
produced by a merger between a large disk with inclination $i_1 =
-109$ and a small disk with inclination $i_2 = 71$; though somewhat
more triaxial than most of the unequal-mass remnants (see
Fig.~\ref{shapes}), it serves to illustrate the kinematic structure of
typical 3:1 merger remnants.  Velocity profiles at selected positions
are shown in Fig.~\ref{lsvdB1}.

The mean velocities along the major axis reveal a rotation curve
similar to those seen in early type spiral and S0 galaxies; the
rotation velocity smoothly increases with distance from the center,
then levels off at larger radii.  The amplitude of this curve is
nearly twice that of remnant~E (Fig.~\ref{paramE}).  As the latter is
the fastest rotator among our sample of equal-mass mergers, it is at
once evident that the 3:1 remnants are kinematically distinct from
their 1:1 counterparts.

Along the major axis, the velocity dispersion declines only gradually
out to about $1.5 R_{\rm e}$, then falls off rapidly at larger radii.
As Fig.~\ref{lsvdB1}a shows, the high dispersion at small radii is due
to a population of box orbits associated with a central bar or
triaxial bulge.  At larger radii the dominant orbit family shifts from
boxes to Z-tubes, most of which rotate in the same direction as the
initial disk of the larger progenitor.  The relatively low dispersions
seen in the outer regions indicate that this disk has survived the
merger without a great deal of dynamical heating.

The 3:1 merger remnants often have rather complex major-axis $h_3$
curves, and remnant~B$_1$ is no exception.  Within $|X| < 0.05$ the
$h_3$ parameter has the same sign as $v_0$; the velocity profile has
broad leading and narrow trailing wings, much as in typical 1:1
remnants.  But at slightly larger $|X|$ values $h_3$ abruptly changes
sign and the shape of the profile is inverted.  Representative
profiles at $X = 0.03$ and $X = 0.09$ are shown in Figs.~\ref{lsvdB1}a
and~\ref{lsvdB1}b, respectively.  As noted above, the former is
dominated by box orbits; its broad leading wing is populated with
Z-tube orbits.  In contrast, the latter is dominated by Z-tubes, and
its broad trailing wing is largely populated by boxes.

\begin{figure}
\begin{center}
\epsfig{figure=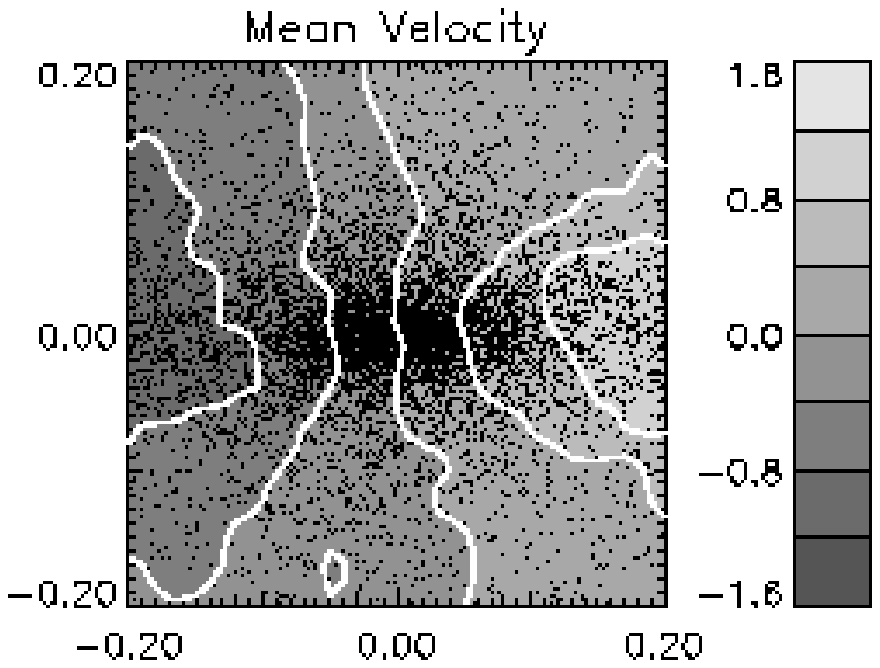,width=2.75in}
\end{center}
\begin{center}
\epsfig{figure=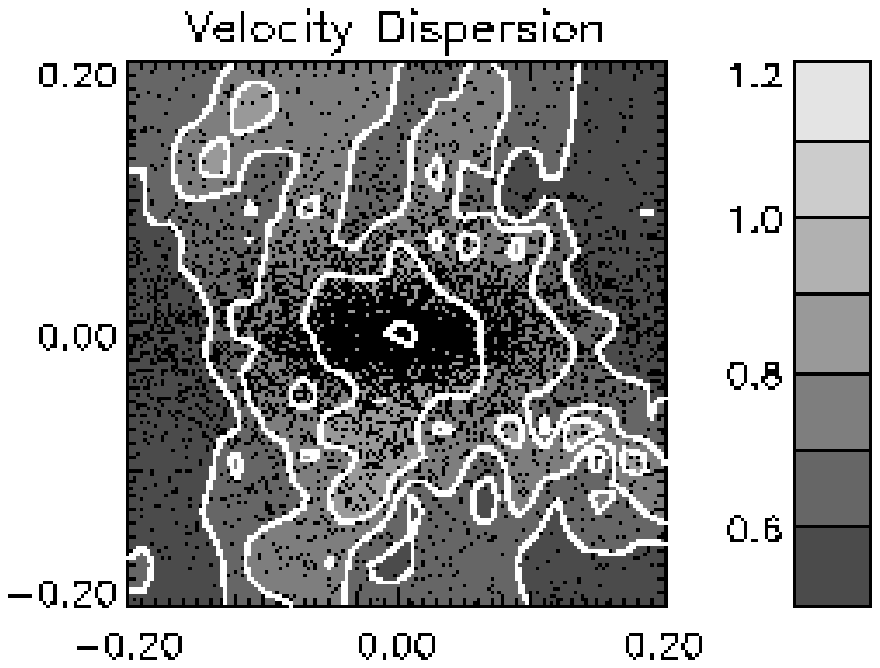,width=2.75in}
\end{center}
\begin{center}
\epsfig{figure=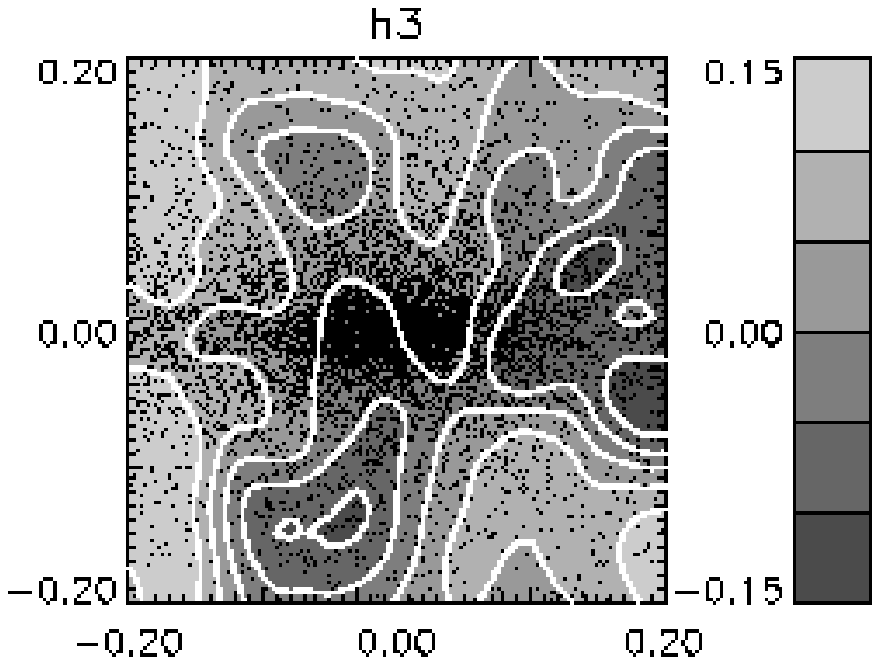,width=2.75in}
\end{center}
\begin{center}
\epsfig{figure=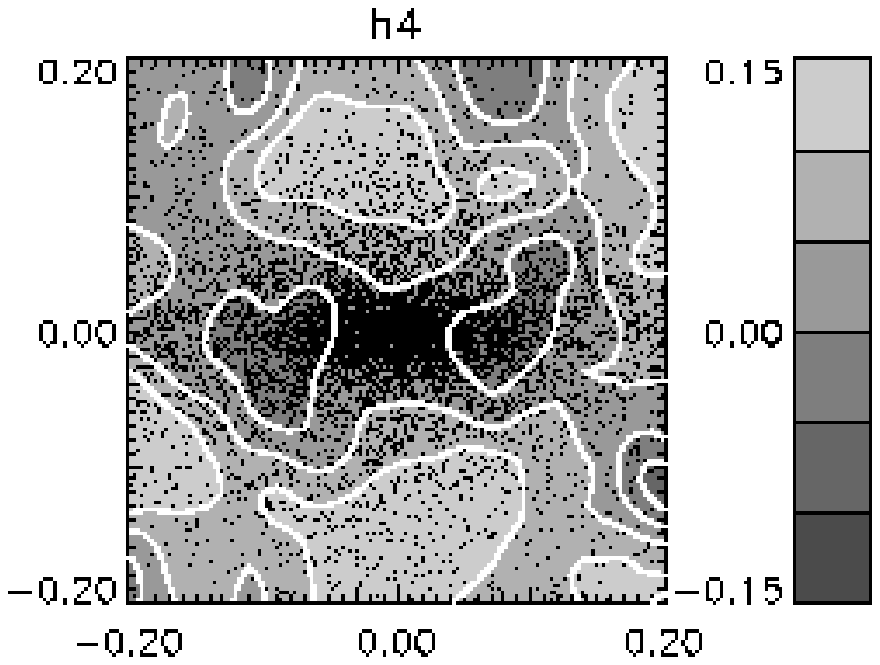,width=2.75in}
\end{center}
\caption{Contour plots of Gauss-Hermite parameters for remnant~B$_1$.}
\label{mapB1}
\end{figure}

Compared to the other cases presented here, remnant~B$_1$ has rather
small $h_4$ values over most of the major axis; while the central peak
in Fig.~\ref{paramB1} is significant, the measured values rapidly fall
to near zero further from the center.  There is some hint that $h_4$
actually becomes negative at intermediate $|X|$, but this is not
compelling as most points have error bars consistent with $h_4 = 0$.
The three rightmost points have positive $h_4$ values which appear
significant; these may be due to incomplete phase mixing as no
corresponding upturn is seen on the other side.

Fig.~\ref{mapB1} presents maps of all four Gauss-Hermite parameters
for remnant~B$_1$.  As a group, the 3:1 mergers have fairly regular
velocity fields; remnant~B$_1$ is typical in this regard, showing
somewhat faster rotation near the disk plane and a nearly cylindrical
rotation pattern elsewhere.  Some asymmetry which may be due to a
long-lived warp is evident, but the zero-velocity contour falls close
to the minor axis, so there is little kinematic misalignment.  The
dispersion contours are somewhat elongated parallel to the minor axis,
but $\sigma$ falls off monotonically with increasing $|Z|$, showing no
sign of the off-axis peaks noted in the 1:1 remnants.  Both the good
kinematic alignment and the lack of off-axis dispersion peaks are
expected in view of the relative scarcity of X-tube orbits in this and
most other 3:1 merger remnants.

The $h_3$ and $h_4$ parameter maps show definite large-scale patterns
which, however, are not easy to interpret.  Particularly puzzling is
the $h_3$ map; $h_3$ is basically inversion-symmetric along the major
axis (Fig.~\ref{paramB1}), but no simple symmetry is seen across the
face of the system.  The $h_4$ map shows peaks on the minor axis above
and below the center of the galaxy.  As Fig.~\ref{lsvdB1}c shows, the
broad-winged profile at these locations is largely due to Z-tube
orbits, with some contribution from X-tubes and boxes.  Curiously,
each wing is dominated by particles from a different progenitor.

\subsection{Prolate 3:1 mergers}

While most of the unequal-mass merger remnants in our sample are much
like the one just described, two have kinematic properties somewhat
reminisicent of the equal-mass remnants.  These are remnants~A$_1$
and~A$_2$, which -- probably not by coincidence -- are also the two
most prolate of the 3:1 remnants (Fig.~\ref{shapes}).  Both objects
were produced by direct encounters between disks with inclinations of
$0$ and $71$; here we describe remnant~A$_1$, which results when the
larger disk has inclination $i_1 = 0$.

\begin{figure}
\begin{center}
\epsfig{figure=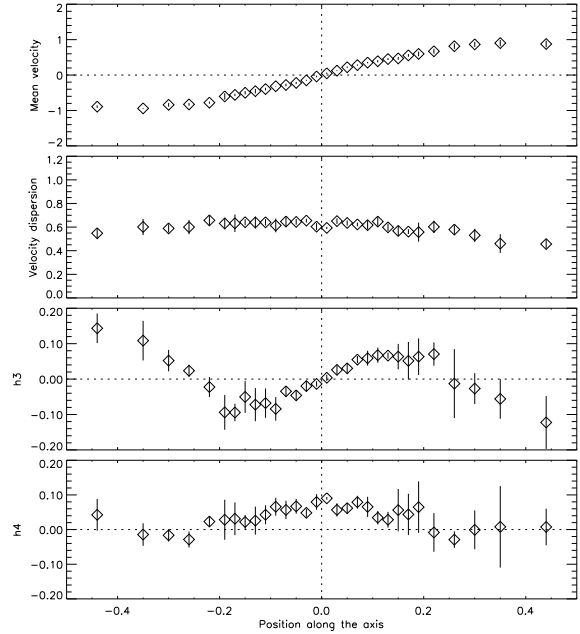,width=3.25in}
\end{center}
\caption{Parameters along the major axis for remnant~A$_1$, a triaxial
3:1 merger.}
\label{paramA1}
\end{figure}

Fig.~\ref{paramA1} shows how the velocity distribution parameters vary
along the major axis of remnant~A$_1$.  While this remnant rotates
faster than any of the 1:1 remnants, its rotation curve rises rather
gradually compared to those of typical 3:1 remnants.  Moreover, the
dispersion profile is nearly flat instead of falling off at large
radii.  These kinematic properties indicate that the larger disk,
while only slightly thickened by the in-plane merger, has been
significantly heated in the radial and azimuthal directions.  In fact,
the encounter triggers the formation of a very strong bar in the
larger disk, and this bar in turn accounts for the nearly-prolate
figure of the final merger remnant.

The major-axis $h_3$ and $h_4$ curves for this object also resemble
those seen in many 1:1 merger remnants.  Over most of the measured
range, $h_3$ has the same sign as $v_0$, indicating that the velocity
profile has a broad leading and narrow trailing wings; only beyond
$\sim 2 R_{\rm e}$ does the profile revert to the shape characterisitc
of a rotating disk.  The $h_4$ parameter is also distinctly greater
than zero along most of the major axis, like many 1:1 remnants but
unlike the oblate 3:1 sample.

\begin{figure}
\begin{center}
\epsfig{figure=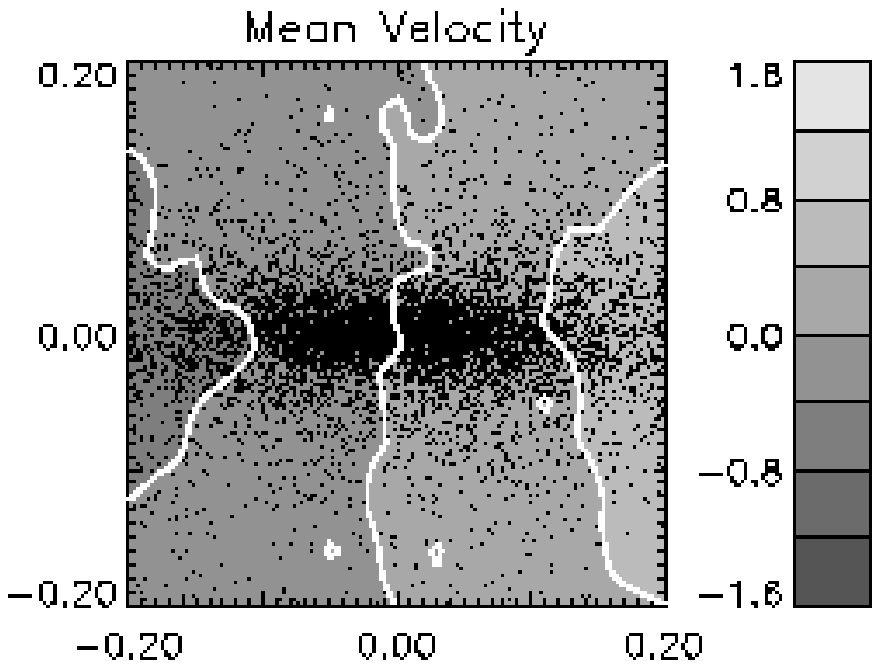,width=2.75in}
\end{center}
\begin{center}
\epsfig{figure=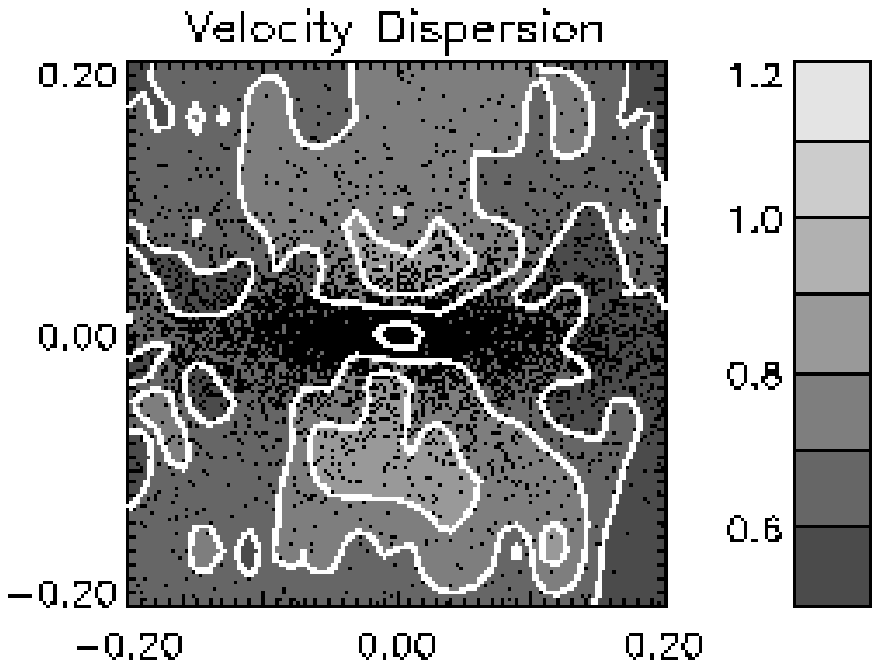,width=2.75in}
\end{center}
\caption{Contour plots of the first two Gauss-Hermite parameters for
remnant~A$_1$.}
\label{mapA1}
\end{figure}

The mean velocity and dispersion maps in Fig.~\ref{mapA1} reveal
kinematic properties intermediate between equal-mass and unequal-mass
merger remnants.  The velocity field shows no sign of kinematic
misalignment; this is typical of 3:1 remnants.  On the other hand, the
dispersion has maxima on the minor axis above and below the center,
like those seen in remnants~G and~C.  Consistent with the explanation
for these peaks advanced in \S~2.2.3, we note that remnant~A$_1$ has a
relatively large population of X-tube orbits for a 3:1 merger.

\section{Conclusions}

During a merger, stellar orbits are scattered by the fluctuating
gravitational potential.  However, the potential settles down long
before orbits can be completely randomized; consequently, merger
remnants preserve significant ``memories'' of their progenitors
(eg.~Barnes 1998 and references therein).  In this study we have shown
that such memories can be partly recovered from the line of sight
velocity profiles of merger remnants.

\subsection{Comparison with observations}

Observations of early-type galaxies reveal a wide variety of kinematic
phenomena similar to those seen in our sample of remnants.  Very
briefly, we will touch on some of these similarities.

\subsubsection{Misaligned rotation}

As pointed out in sections 2.1 and 2.2.3 as well as in previous
studies, kinematic misalignments are expected in merger remnants, and
especially in equal-mass mergers.  Franx, Illingworth, and de Zeeuw
(1991) present a study of kinematic misalignment in elliptical
galaxies; most of the galaxies they observed have small misalignments.
While remnant~C (Fig.~\ref{mapC}) is dramatically misaligned, as a
whole the equal-mass remnants described here are better aligned than
samples reported in earlier work (Barnes 1992).  The incidence of
severe misalignment probably depend on several factors; for example,
central density profile can have a significant impact on the
phase-space available to major-axis tube orbits.  Until the factors
which favor misalignment are better understood, it's not clear if the
observed scarcity of severe kinematic misalignment can constrain the
role of equal-mass mergers in the formation of elliptical galaxies.

\subsubsection{Kinematically decoupled cores}

Hernquist and Barnes (1991) presented a dissipational simulation
showing that the core of a merger remnant could decouple and
counterrotate.  We have examined the quantitative effect that
counterrotation can have on the observed kinematics of merger
remnants.  Several galaxies, such as NGC 1700 (Statler, Smecker-Hane,
\& Cecil 1996), NGC 4365, NGC 4406, NGC 5322 (Bender \& Surma 1992),
IC 1459, NGC 1374, NGC 4278 (van der Marel \& Franx 1993), NGC 4816,
and IC 4051 (Mehlert et al. 1998) all show line of sight kinematics
similar to the line of sight velocity distributions of our models
(though some of these galaxies are strong candidates for other
scenarios that create counterrotation).  In particular, we find
amplitudes of $h_3$ and $h_4$ similat to those reported in the
observational studies.  This shows that major merger can produce
remnants with the degree of skewness and kurtosis observed in
counterrotating systems.  We expect that as more galaxies are observed
further examples with line of sight velocity distributions similar to
ours will be found.

Also worth noting are the observations of NGC 253 by Anantharamaiah \&
Goss (1996) which found an orthogonally rotating core which was
suspected to be caused by a merger event.  One of our 1:1 merger
models also produced an orthogonally rotating core (see
Fig.~\ref{mapG}).

\subsubsection{Counterrotating populations}

Early-type galaxies with extended counterrotating populations are rare
but not unknown.  Some of these systems may have formed by episodic
galaxy building (Thakar \& Ryden 1996), but others are harder to
explain in this way.  For example, NGC~4550 (Rubin et al.~1992) has
counterrotating disks of comparable radial extent and luminosity;
Pfenniger (1999) has proposed this galaxy formed by an in-plane merger
of two disk galaxies.  Our analysis of remnant~H shows that a somewhat
wider range of merger scenarios can produce counterrotating
populations.

\subsubsection{Rapid rotators}

Barnes and Hernquist (1992) and Schweizer and Seitzer (1992), among
others, have suggested that S0 galaxies could be made by mergers.
Fisher (1997) has collected a sample of S0 galaxies with line of sight
velocity distributions fit using Gauss-Hermite parameters.  Comparing
his observations to our simulations, we find a good match between
Fisher's parameters and the parameters for our disky 3:1 mergers.  The
overall shapes of the Gauss-Hermite parameters plotted on the major
axis are remarkably similar, except that our rotation curve near the
origin is less steep than the observed S0 galaxies, and some details
near the center of our simulations (such as the first twist in the
$h_3$ parameters) are not apparent in Fisher's data.

\begin{figure}
\begin{center}
\epsfig{figure=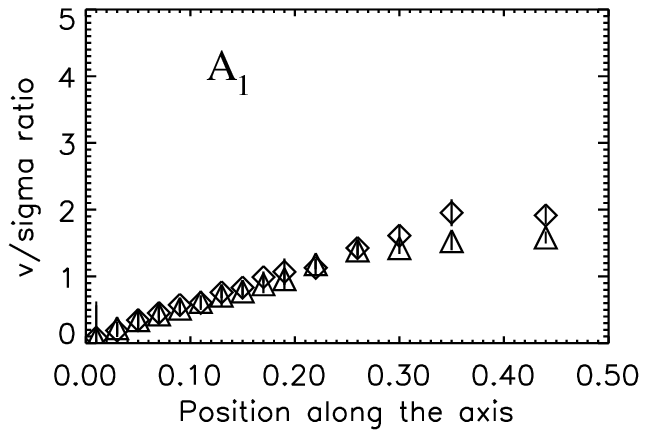,width=1.6in,bbllx=26,bblly=6,bburx=210,bbury=129}
\epsfig{figure=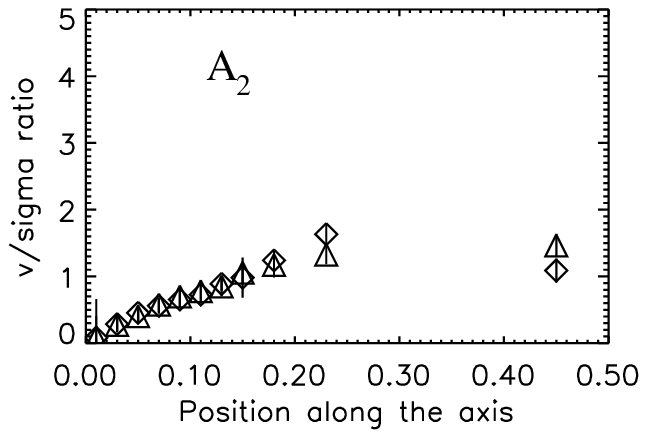,width=1.6in,bbllx=26,bblly=6,bburx=210,bbury=129}
\end{center}
\begin{center}
\epsfig{figure=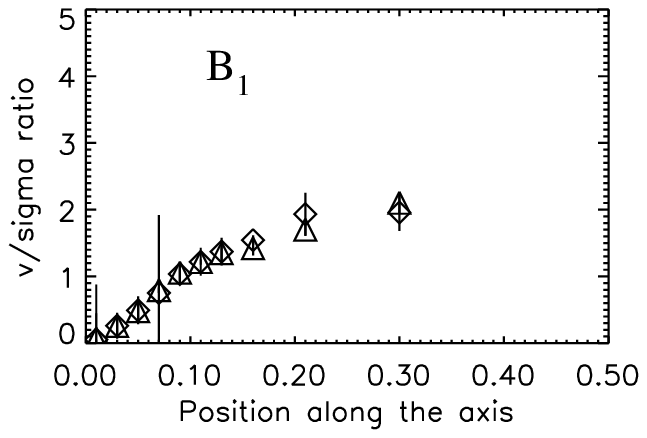,width=1.6in,bbllx=26,bblly=6,bburx=210,bbury=129}
\epsfig{figure=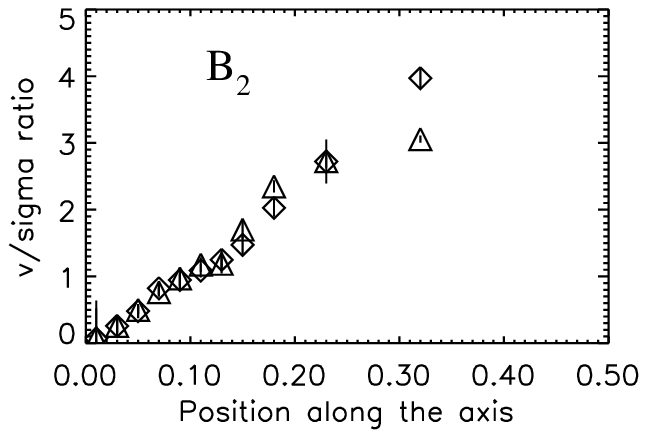,width=1.6in,bbllx=26,bblly=6,bburx=210,bbury=129}
\end{center}
\begin{center}
\epsfig{figure=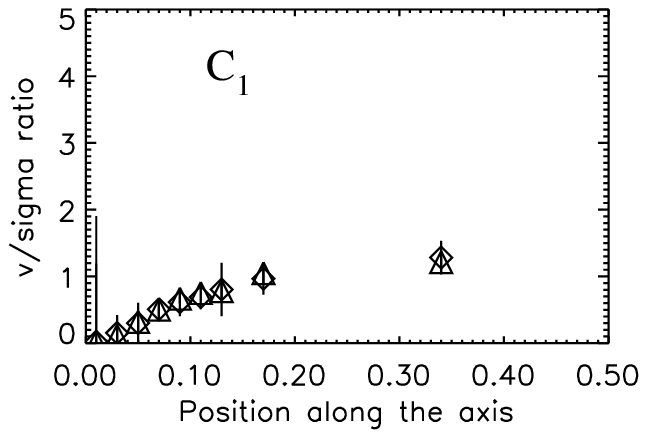,width=1.6in,bbllx=26,bblly=6,bburx=210,bbury=129}
\epsfig{figure=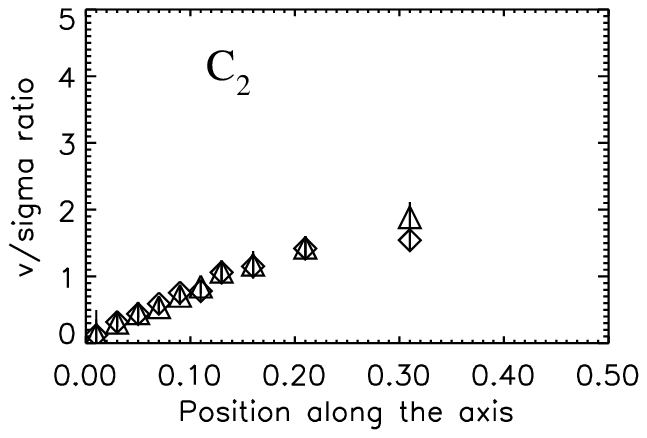,width=1.6in,bbllx=26,bblly=6,bburx=210,bbury=129}
\end{center}
\begin{center}
\epsfig{figure=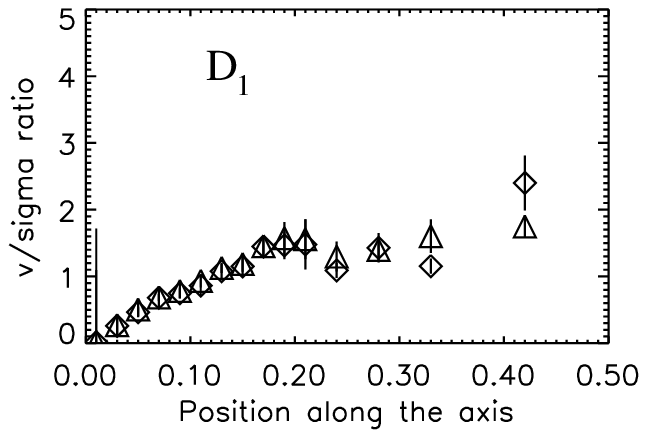,width=1.6in,bbllx=26,bblly=6,bburx=210,bbury=129}
\epsfig{figure=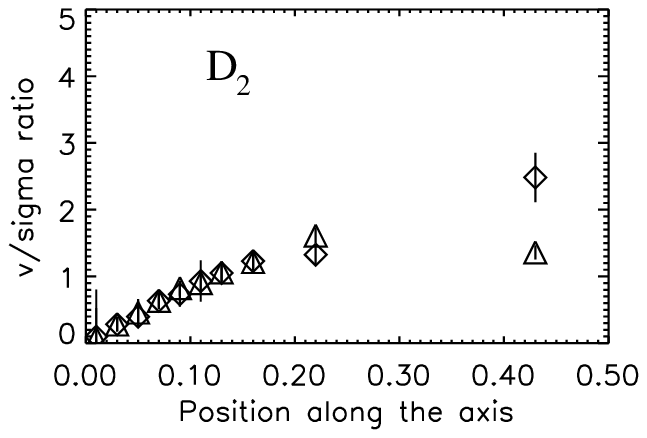,width=1.6in,bbllx=26,bblly=6,bburx=210,bbury=129}
\end{center}
\caption{Ratio of $v_0 / \sigma$ as a function of major-axis position
for all 3:1 remnants.  Different plotting symbols refer to opposite
sides of the major axis.}
\label{vsigma}
\end{figure}

Based on measurements of the ratio of mean velocity to velocity
dispersion for a set of faint elliptical galaxies, Rix, Carollo, and
Freeman (1999) have argued that these galaxies rotate too rapidly to
be products of dissipationless mergers.  When we compare these
measurements to $v/\sigma$ ratios we measured in our simulations
(Fig.~\ref{vsigma}), we find that unequal-mass mergers can not only
produce the same peak $v/\sigma$ ratios but also produce the same
relations between $v/\sigma$ and radius.  Moreover, both our results
and the results of Rix, Carollo, and Freeman show similar ranges of
maximum $v/\sigma$ ratios, with values in the range of $\sim 1$ to
$\sim 4$.  We conclude that unequal mass mergers can produce remnants
with the dynamics, including the $v/\sigma$ ratios, characteristic of
these faint ellipticals.  This result complements a recent study by
Naab, Burkert, \& Hernquist (1999), which finds that unequal-mass
mergers can also produce the disky isophotes characteristic of faint
ellipticals and S0 galaxies.

\subsection{Summary}

We have used Gauss-Hermite expansions to measure the line of sight
velocity profiles of simulated merger remnants.  Even relatively
modest values of $N$ provide enough data to obtain significant
detections of non-Gaussian profiles.  Some key results are listed
below.

1. Equal-mass merger remnants exhibit a variety of kinematic features
rather than any single unique ``merger signature''.  However, certain
features seem common to most of the remnants in our sample; these
include slowly rotating inner regions, relatively flat dispersion
profiles, off-axis dispersion peaks, and velocity distributions with
broad leading and narrow trailing wings.

2. Unequal-mass merger remnants show much less variation in kinematic
properties; instead, the larger disk often survives with only moderate
damage.  Such disk-dominated remnants are characterized by relatively
rapid rotation, falling dispersion profiles, and velocity
distributions with narrow leading and broad trailing wings.  For mass
ratios of 3:1, between half and three-fourths of the remnants in our
study had strong disk-like kinematics.

3. Simulated remnants have many kinematic characteristics similar to
those observed in early-type galaxies.  For example, we described
counterrotating populations, misaligned rotation, and kinematically
decoupled cores resembling those reported in some elliptical galaxies,
and rapid rotation consistent with faint ellipticals and S0 galaxies.
However, our simulations don't always match observed galaxies.  For
example, the mean velocity and $h_3$ parameters usually have opposite
signs in luminous elliptical galaxies (eg.~Bender, Saglia, \& Gerhard
1994), while these parameters often have the same sign in our
simulated remnants.  More work needs to be done to examine the
connections between simulated remnants and real galaxies; in
particular, the effects of random viewing angles must be taken into
account before definitive comparisons of models and observations are
possible.

We thank Hans-Walter Rix and Andreas Burkert for stimulating
discussions, and the referee for a prompt and helpful report.  JEB
acknowledges partial support from NASA grant NAG 5-8393.


\end{document}